\documentclass[aps,prd,twocolumn,superscriptaddress,nofootinbib,
  floatfix,longbibliography]{revtex4-2}

\usepackage{graphicx}
\usepackage{amsmath,amssymb}
\usepackage{booktabs}
\usepackage{xcolor}
\usepackage{tikz}
\usetikzlibrary{arrows.meta,positioning,fit}
\usepackage[colorlinks=true,linkcolor=blue!60!black,citecolor=blue!60!black,urlcolor=blue!60!black]{hyperref}

\graphicspath{{figures_v7/}{figures/}}

\newcommand{\pt}{\ensuremath{p_{\mathrm{T}}}}
\newcommand{\HT}{\ensuremath{H_{\mathrm{T}}}}

\newcommand{\deta}{\ensuremath{\Delta\eta}}

\newcommand{\dptr}{\ensuremath{\Delta\pt/\pt}}
\newcommand{\eeZ}{\ensuremath{e^+e^-\to Z\to q\bar{q}}}

\begin{document}

\title{Cross-Detector Transfer Learning with Parnassus: From ALEPH to SLD}

\author{Ya-Feng Lo}
\email{ya-feng.lo@cern.ch}
\affiliation{Department of Particle Physics and Astrophysics, Stanford University, Stanford, CA 94305, USA}
\affiliation{Fundamental Physics Directorate, SLAC National Accelerator Laboratory, Menlo Park, CA 94025, USA}

\author{Chi Lung Cheng}
\email{alkaidc@stanford.edu}
\affiliation{Department of Particle Physics and Astrophysics, Stanford University, Stanford, CA 94305, USA}
\affiliation{Fundamental Physics Directorate, SLAC National Accelerator Laboratory, Menlo Park, CA 94025, USA}

\author{Benjamin Nachman}
\email{nachman@stanford.edu}
\affiliation{Department of Particle Physics and Astrophysics, Stanford University, Stanford, CA 94305, USA}
\affiliation{Fundamental Physics Directorate, SLAC National Accelerator Laboratory, Menlo Park, CA 94025, USA}

\date{\today}

\begin{abstract}
Parnassus is a fast detector-simulation and reconstruction framework that maps truth-level particles directly to reconstructed particles. In this work, we investigate cross-detector transfer learning by adapting a Parnassus model trained on the ALEPH detector to the SLD detector. ALEPH and SLD have a similar detector response, as they were both targeting hadronic $Z$-pole $e^+e^-$ events, while differing substantially in the underlying detector technology, reconstruction, and archived data representation. We initialize the SLD particle model with weights learned on ALEPH, fine-tune it on SLD, and compare it with the same architecture trained directly on SLD. The ALEPH-initialized model gives substantially improved particle- and jet-level agreement with the SLD reference, including a factor of $5.2$ improvement in the charged-particle angular response and an improvement in the jet angular resolution from $1.73\times$ to $1.15\times$ the SLD reference. The results show that detector-response information learned on ALEPH can be reused when modeling SLD and motivate reusable pretrained Parnassus models for legacy $e^+e^-$ detectors, which is especially critical for experiments without access to the original software pipeline.  With this paper, we also release an AI-ready version of simulated SLD $e^+e^-$ events.
\end{abstract}

\maketitle

\section{Introduction}
\label{sec:intro}

Fast and accurate detector simulation is an essential part of experimental
particle physics. Parnassus~\cite{Dreyer:2024bhb,Dreyer:2025events,
Elabd:2026parnassus,Elabd:2026differentiable}
is a framework for automatically tuning the map from
truth-level particles to reconstructed objects. It includes both parametric
and deep-generative approaches. In the generative approach used here,
given the truth-level particles in an event, the model generates the
corresponding reconstructed particles and event-level quantities.
In addition to generative models, Parnassus supports
differentiable parametric, Delphes-like detector-response models whose
parameters can be optimized directly using simulated detector data.
Parnassus has been applied to proton-proton collision such as CMS Open
Data~\cite{Dreyer:2024bhb,Dreyer:2025events},
 fast simulation and reconstruction in
ATLAS~\cite{ATLAS:2026parnassus},
and, more recently, to electron-positron collision, for instance, archived ALEPH simulated events at
LEP~\cite{Lo:2026aleph} as well as the CLD detector concept for the FCC-ee. In the ALEPH study, the model was shown to reproduce
the energy-flow reconstruction for a range of event-, jet-, and particle-level
observables.

For preserved collider experiments, the available simulated samples are
limited to those that remain in the archive, and the original simulation and
reconstruction software is not always possible or practical to run in a
modern computing environment. This motivates the use of existing simulated
samples as efficiently as possible. In particular, a model already trained
for one experiment may provide a useful starting point for another experiment
with a related physics process.
Pretraining and fine-tuning have provided reusable
representations for generative and other downstream tasks in collider
physics~\cite{Mikuni:2025omni}, while cross-detector transfer has been
demonstrated for machine-learned particle-flow
reconstruction~\cite{Mokhtar:2025pflowtransfer}.

We study this possibility with ALEPH and SLD. Both experiments recorded
hadronic $Z$-pole events at $\sqrt{s}\simeq 91.2$\,GeV. Thus, even though
their detector and reconstruction systems are different, the detectors
were designed to operate in a similar kinematic regime.
ALEPH operated at LEP and SLD at the SLC, with different tracking,
calorimetry, and particle-identification detectors. SLD also operated with a
longitudinally polarized electron beam, which makes its preserved $Z$-pole
dataset particularly valuable for studies of polarization-dependent
observables. A modern detector-simulation model for SLD would therefore
support both the preservation and the future reanalysis of this dataset.


The archived samples used here also have different reconstruction conventions
and data formats. The ALEPH sample is obtained from a modern EDM4hep
conversion of the archived ALEPH simulation~\cite{Defranchis:2026aleph}.
The SLD sample is constructed from archived Jazelle mini-DST Monte Carlo
files translated into a modern columnar
format~\cite{Cheng:2026sld}. The detector and data samples are described in
more detail in Sections~\ref{sec:detectors} and~\ref{sec:slddata}.

For SLD, we compare two training procedures using the same Parnassus
architecture and the same SLD training sample. The first model is initialized
randomly and trained on SLD. We refer to this as the
\emph{directly trained SLD model}. The second model starts from the
particle-network parameters of a Parnassus model trained on ALEPH and is then
fine-tuned on SLD. Apart from the initialization, the training configuration
is kept the same for the two models.

The ALEPH-initialized model gives better agreement with the SLD reference for
particle- and jet-level quantities, while the event-level performance of the
two models is similar. The ALEPH initialization therefore provides a useful
starting point for modeling the SLD detector despite the differences between
the two detector and reconstruction systems.

Section~\ref{sec:detectors} describes the ALEPH and SLD detector and data
domains. Section~\ref{sec:slddata} describes the construction of the SLD
Parnassus training sample from the archived mini-DSTs.
Section~\ref{sec:domains} compares the ALEPH and SLD quantities used for
training. Section~\ref{sec:model} describes the Parnassus model and the
fine-tuning procedure. Section~\ref{sec:results} presents the results.

\section{The ALEPH and SLD detector and data domains}
\label{sec:detectors}

\subsection{SLD at the SLC}
\label{sec:sld}

The SLAC Linear Collider (SLC) was a linear $e^+e^-$ collider operating at
the $Z$ pole from 1989 to 1998~
\cite{SLC:1984design,Cheng:2026sld}.
The SLD experiment collected data from 1992 to
1998\textcolor{blue}{~\cite{SLD:1984design}}. A distinctive feature of the
SLC program was the longitudinally polarized electron beam, with an average
polarization of approximately 73--77\% during the 1996--1998 running
period. The polarized data
enabled precision electroweak measurements, including the determination of
the weak mixing angle from the left-right asymmetry
$A_{LR}$~\cite{SLD:2000alr}. The preserved SLD dataset therefore provides
access to polarization-dependent observables that are not available in the
other $Z$-pole collider datasets. A modern detector simulation for SLD would
support future studies using these data.

The SLC Large Detector (SLD)~\cite{SLD:1984design} surrounded the interaction
point with several detector systems:
\begin{itemize}
  \item \textbf{VXD3}, a CCD vertex detector used for precise vertex and
    impact-parameter measurements~\cite{Abe:1997vxd3};
  \item the \textbf{Central Drift Chamber (CDC)}, providing
    charged-particle tracking and ionization measurements ($dE/dx$);
  \item the \textbf{Cherenkov Ring Imaging Detector (CRID)}, providing
    particle-identification information;
  \item the \textbf{Liquid Argon Calorimeter (LAC)}, measuring
    electromagnetic and hadronic energy deposits;
  \item the \textbf{Warm Iron Calorimeter (WIC)}, providing muon
    identification and serving as the magnetic flux return.
\end{itemize}

Measurements from these subsystems were processed into
several classes of analysis-level reconstructed objects. Reconstructed
charged tracks include helix parameters and covariance matrices, together
with hit counts, fit-quality information, and ionization measurements from
the CDC and VXD3. Calorimeter information is stored as clusters containing
total and layer-by-layer energy deposits from the LAC. Additional records
contain particle-identification likelihoods from the CRID, muon-track
segments from the WIC, and electron--calorimeter matching information.
Relational tables preserve associations between objects reconstructed in
different subsystems, such as charged tracks and calorimeter
clusters. Higher-level particle summaries produced by
the SLD reconstruction provide the reconstructed-particle targets used to
train Parnassus in this study.

SLD data were stored using the Jazelle data-management and I/O system, which
was developed for the experiment's Fortran and Mortran software environment.
The analysis-level event records are stored as mini Data Summary Tapes
(mini-DSTs), organized as collections of banks containing reconstructed
tracks, calorimeter information, particle summaries, and other event
quantities. Simulated mini-DSTs additionally contain generator and simulation
information.

A recent SLD data-preservation effort developed the open-source
\texttt{jazelle} package to read the original binary files and translate
their bank structure into modern array-based
formats~\cite{Cheng:2026jazelle}. The same effort also digitized
SLD internal documentation that is used in this work to interpret the
archived bank structure. The present study uses archived SLD Monte Carlo
mini-DSTs translated with \texttt{jazelle}. The extraction of truth and
reconstructed particles from these files and their conversion to the
Parnassus training representation are described in
Section~\ref{sec:slddata}.

\subsection{ALEPH at LEP}
\label{sec:aleph}

ALEPH~\cite{ALEPH:1990ay,ALEPH:1994ay} was one of the four experiments at the Large Electron--Positron Collider (LEP) at CERN. The detector included a time-projection chamber for charged-particle tracking, electromagnetic and hadronic calorimeters, and an energy-flow reconstruction that combined information from the tracking and calorimeter systems.

The ALEPH sample used in this study consists of simulated hadronic $Z$ decays at $\sqrt{s}\simeq 91.2$\,GeV. The events were processed with the \textsc{Geant}3-based ALEPH detector simulation ~\cite{Brun:1987ma} and the ALEPH energy-flow reconstruction~\cite{ALEPH:1994ay}. The archived simulated samples were subsequently converted to EDM4hep as part of a recent reinterpretation effort ~\cite{Defranchis:2026aleph}.  Approximately $1.0$ million events are used to train the ALEPH Parnassus model.

The ALEPH Parnassus model was previously studied and validated at the event, jet, and particle levels~\cite{Lo:2026aleph}. In the present study, the particle-network parameters learned from the ALEPH sample are used to initialize the SLD model before training on the SLD sample.

\subsection{ALEPH and SLD Detector Domains}

The two experiments provide a controlled setting for studying transfer. The truth-level inputs are statistically similar hadronic $Z$ decays, while the mapping from truth particles to reconstructed particles differs because of the detector response, reconstruction algorithms, particle-class definitions, kinematic acceptance of the archived formats, and available sample sizes.  Section~\ref{sec:domains} compares these differences using the actual training inputs.

\section{Construction of the SLD Parnassus dataset}
\label{sec:slddata}

The SLD sample used in this study consists of nine archived Monte Carlo mini-DST files containing 260{,}948 events. The files were translated from the original Jazelle bank format to Parquet using the \texttt{jazelle} tools described in Refs.~\cite{Cheng:2026sld,Cheng:2026jazelle}. Truth-particle information is obtained from the \texttt{MCPART} bank and reconstructed-particle information from the \texttt{PHPSUM} bank. The information extracted from these banks is summarized in Table~\ref{tab:mapping}. During the development of these tools, large language models
were used to assist with recovering bank layouts that were incompletely
documented or unsupported by the surviving reader software. Candidate field
positions and data types inferred from the archived documentation and
recurring binary patterns were validated using physics constraints and their
consistency across multiple events~\cite{Cheng:2026sld}.

\subsection{Particle information}

The \texttt{MCPART} bank contains generator-level and \textsc{Geant}-generated particles with four-momentum $(p_x,p_y,p_z,E)$, charge, an internal particle-type code \texttt{ptype}, parent information, position information, and an \texttt{origin} field.  Since the bank contains both particles from the generated event and particles produced during detector simulation, the truth-particle selection uses the particle-type and origin information to retain detector-stable particles while removing \textsc{Geant} secondaries.

The \texttt{ptype} field is a legacy SLD particle code rather than a PDG identifier. We construct the \texttt{ptype}-to-PDG mapping by identifying each code from the generator record. The stored energies are not used for this purpose: they are reconstructed from an internal mass table, so the mass inferred from the stored four-momentum reproduces that table rather than providing an independent identification. The identification instead uses the production rate of each code as a function of the primary quark flavor, the species of its parent, the particles produced together with it, and energy--momentum conservation in the decays in which it appears. On this basis, codes 7--10 are identified as $e$, $\nu_e$, $\mu$, and $\nu_\mu$: they are produced almost exclusively in decays of charmed and bottom hadrons, each neutrino code appears only together with the charged lepton of the same family, and the decays $D^0\to K\ell\nu$ and $\pi^0\to\gamma e^+e^-$ conserve energy only when codes 7 and 9 are assigned the electron and muon masses. The same procedure identifies codes 41 and 42 as protons and neutrons, for which $\Delta(1232)$ and $\Lambda_c$ decays reproduce the parent masses only with the nucleon mass assignment. A PDG identifier is assigned only for codes identified in this way; otherwise it is set to zero. Electron and muon truth labels are assigned from codes 7 and 9.

Neutrinos are removed from the truth-particle selection, following the convention used for the ALEPH sample. The generator record also represents the decay of a $K^0$ into $K^0_S$ or $K^0_L$, and of a $\Sigma^0$ into $\Lambda\gamma$, by storing both the parent and its daughters with identical momenta. The parent entries are removed so that each particle enters the truth sample once.

Truth kinematics are calculated from the stored momenta, and the truth mass is assigned from the identified species. The position fields in \texttt{MCPART} were found to correspond to particle termination points, associated with a decay or interaction, rather than production vertices.  Since no validated production-vertex field was identified in the archived bank, the truth vertex coordinates are set to zero.

The reconstructed particles are obtained from the \texttt{PHPSUM} bank. Each entry contains momentum, charge, and position information for a reconstructed particle summary. Reconstructed kinematics are calculated from the stored momenta. Particle-identification information exists in the archived mini-DSTs, but it could not be reliably associated with the \texttt{PHPSUM} particle summaries used in this study.

We therefore use the common Parnassus class vocabulary while populating only two reconstructed classes for SLD. All charged reconstructed particles are assigned to the charged-hadron class, and all neutral reconstructed particles are assigned to the neutral-hadron class. Only validated position information is retained.

\begin{table*}[t]
\centering
\caption{Information extracted from the SLD mini-DST banks for the
construction of the SLD Parnassus sample.}
\label{tab:mapping}
\small
\begin{tabular}{lll}
\toprule
Quantity & SLD source & Construction \\
\midrule
truth $\pt,\eta,\phi$ &
  \texttt{MCPART} $p_x,p_y,p_z$ &
  calculated from the stored momentum \\
truth mass &
  \texttt{MCPART} \texttt{ptype} &
  assigned from the identified species \\
truth PDG ID &
  \texttt{MCPART} \texttt{ptype}, charge &
  assigned from the species identified in the generator record \\
truth vertex &
  \texttt{MCPART} position &
  set to zero because the stored positions are termination points \\
truth-particle selection &
  \texttt{MCPART} \texttt{ptype}, \texttt{origin} &
detector-stable particles, with simulation secondaries, \\
 & & neutrinos, and duplicated decay-chain entries removed \\
\midrule
reco $\pt,\eta,\phi$ &
  \texttt{PHPSUM} $p_x,p_y,p_z$ &
  calculated from the stored momentum \\
reco position &
  \texttt{PHPSUM} $x,y,z$ &
  validated position information retained \\
reco class &
  \texttt{PHPSUM} charge &
  charged $\to$ charged-hadron class; neutral $\to$ neutral-hadron class \\
\bottomrule
\end{tabular}
\end{table*}

\subsection{Event selection and dataset split}

A small fraction of the archived events contains unphysical reconstructed tracks with very large momenta. Events with reconstructed $\HT=\sum\pt^{\rm reco}\geq200$\,GeV are removed. This selection removes 654 events from the original 250{,}948-event training sample and 28 events from the 10{,}000-event test sample.

After this selection, 250{,}294 events remain for training and validation. Of these, 212{,}500 are used for training and 37{,}794 for validation. The fixed test sample contains 9{,}972 events and is used for all comparisons presented in this paper.

\subsection{Comparison with 1996 data}
\label{sec:datamc}

Since Parnassus is trained to reproduce the archived SLD simulation, we first compare this simulation with recorded SLD data at reconstruction level. We use translated data from the 1996 run and the simulated sample described above, with reconstructed particles taken from the PHPSUM bank. A common hadronic selection is applied to both samples, requiring at least seven charged reconstructed particles, visible energy above 18~GeV, and $H_T<200$~GeV.

\begin{figure*}[t]
\centering
\includegraphics[width=\textwidth]{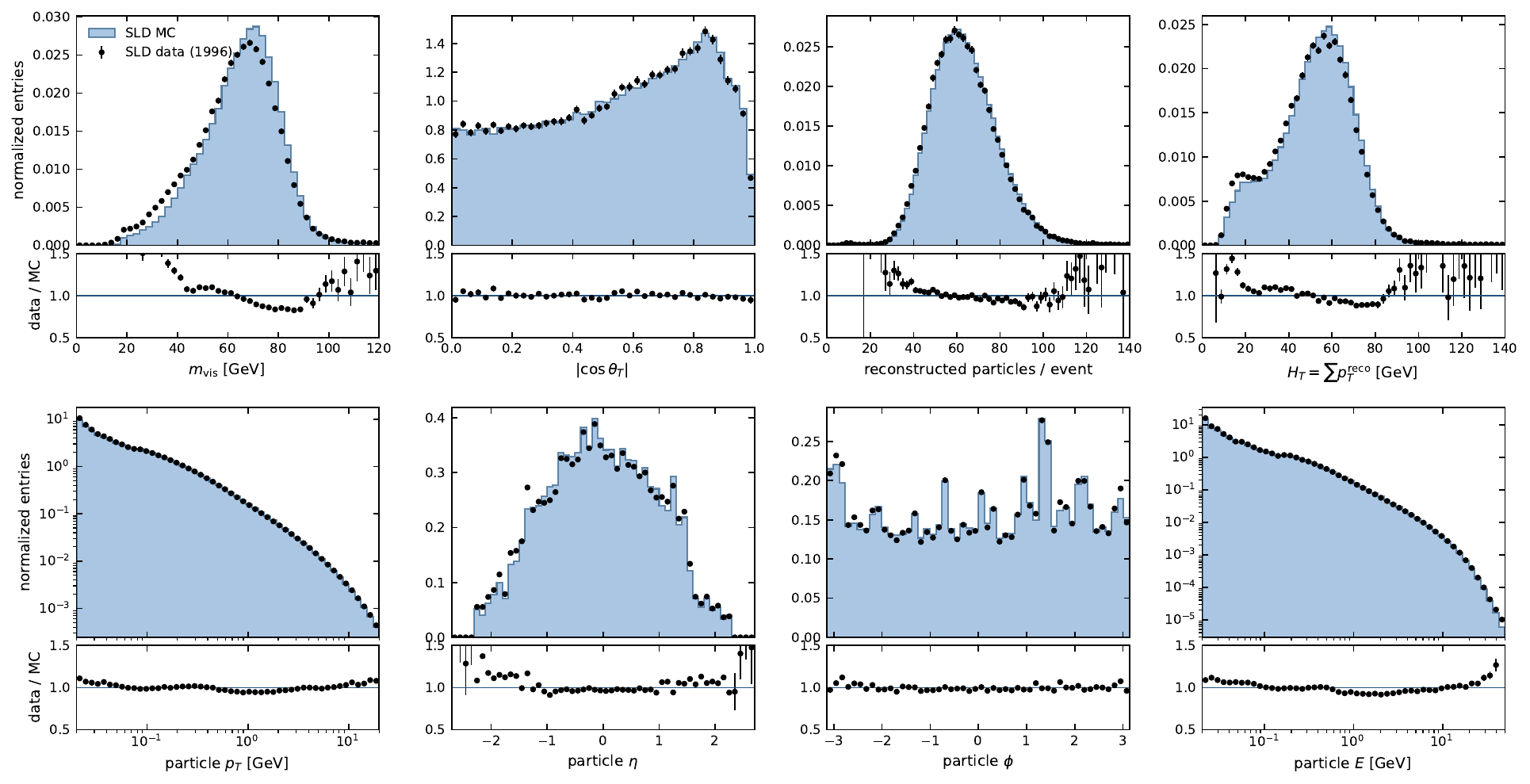}
\caption{SLD 1996 data (points) compared with the simulated sample (histogram) at reconstruction level, with the hadronic selection described in the text applied to both. Top row: visible mass, thrust-axis polar angle, reconstructed particle multiplicity, and \HT. Bottom row: the reconstructed particle four-momentum. All distributions are normalized to unit area, and the lower panels show the data/MC ratio with combined statistical uncertainties. The visible-mass difference is a neutral-energy scale effect: the charged sector agrees to 0.2\%, while each neutral object carries 11\% less energy in the data.}
\label{fig:datamc}
\end{figure*}

Figure~\ref{fig:datamc} compares representative event- and particle-level observables. The archived simulation describes the data well for most distributions, including the thrust-axis polar angle, reconstructed-particle multiplicity, $H_T$, and particle kinematics. The largest difference is seen in the visible-mass distribution. This difference is mainly associated with the neutral-particle response: the charged component is well described, while neutral particles carry less reconstructed energy in the data than in the simulation.

\section{ALEPH vs.\ SLD: the training domains Parnassus sees}
\label{sec:domains}

ALEPH and SLD provide a useful setting for studying cross-detector transfer.  Both samples consist of simulated hadronic $Z$ decays at $\sqrt{s}\simeq91.2$\,GeV, so their truth-level particle distributions are expected to be similar.  The two experiments, however, have different detectors and reconstruction procedures. Their reconstructed-particle distributions therefore characterize the detector-domain shift relevant for the transfer study.

\begin{figure*}[t]
\centering
\includegraphics[width=\textwidth]{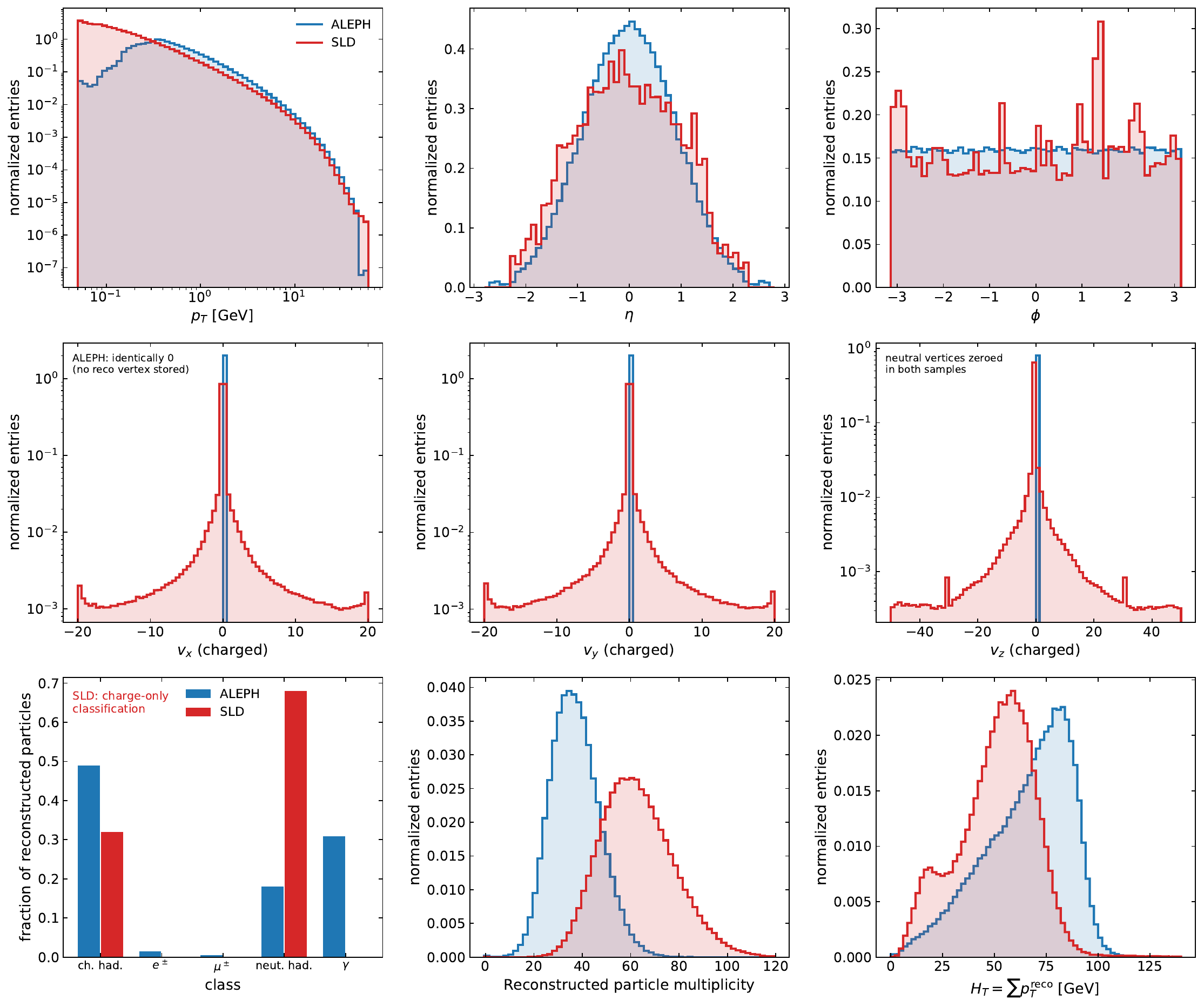}
\caption{Comparison of reconstructed quantities in the ALEPH and SLD training samples. The top row shows particle \pt, $\eta$, and $\phi$. The middle row shows the charged-particle position coordinates $v_x$, $v_y$, and $v_z$. The bottom row shows the reconstructed particle-class composition, particle multiplicity, and event $\HT=\sum\pt^{\rm reco}$. ALEPH populates five reconstructed particle classes, while the SLD representation populates only the charged-hadron and neutral-hadron classes.}
\label{fig:domains}
\end{figure*}

Figure~\ref{fig:domains} compares the reconstructed quantities used in the ALEPH and SLD training samples. The same Parnassus preprocessing is applied to both samples. The figure uses the first 150{,}000 events from each sample.

The reconstructed distributions differ substantially between the two samples.  The median reconstructed-particle \pt\ is 0.87\,GeV for ALEPH and 0.16\,GeV for SLD. The fraction of reconstructed particles below 1\,GeV is 55\% for ALEPH and 82\% for SLD. The mean reconstructed particle multiplicity is 37 for ALEPH and 63 for SLD, while the mean event \HT\ is 66\,GeV and 51\,GeV, respectively.  Charged particles account for 51\% of the ALEPH reconstructed sample and 32\% of the SLD sample.

Differences are also visible in the angular and position distributions. The reconstructed $\phi$ distribution in SLD is non-uniform, and its $\eta$ distribution contains structures near the acceptance boundaries. The corresponding ALEPH distributions are smoother. The available position information also differs between the two archived representations. The ALEPH ntuples used in this study do not contain reconstructed position coordinates for the Parnassus particle features, while position information is available for charged SLD particles.

Some features in Fig.~\ref{fig:domains} also reflect differences in the archived data representation. The particle-class composition is affected by the reduced SLD classification described in Section~\ref{sec:slddata}. The upper end of the SLD \HT\ distribution is affected by the event-cleaning requirement. The absence of reconstructed position information in the ALEPH ntuples is a property of the converted sample rather than of the ALEPH detector.

In contrast, the truth-level particle distributions of the two samples are closely matched. The median truth-particle \pt\ is 0.68\,GeV for both samples, with 41.7\% and 41.6\% of truth particles below 0.5\,GeV for ALEPH and SLD respectively. This agreement is expected because both samples describe the same hadronic $Z$-pole process. The similar truth-level distributions together with the different reconstructed distributions define the detector and reconstruction shift studied in the following sections.

\section{Model and cross-detector fine-tuning}
\label{sec:model}

\subsection{Parnassus representation}

Parnassus models reconstructed events conditionally on the corresponding truth-level particles. The framework contains an event network and a particle network.  The event network generates global reconstructed quantities from the truth event, and these quantities are subsequently used to condition the particle network.

Each particle is represented by
\[
(\pt,\eta,\phi,v_x,v_y,v_z,c),
\]
where $c$ denotes the particle class. The common Parnassus class vocabulary used in this study contains five categories: charged hadrons, electrons, muons, neutral hadrons, and photons. All five classes are populated in the ALEPH sample.  For SLD, the particle-identification information available in the archived particle summaries does not support the same separation. The reconstructed SLD sample therefore populates only the charged-hadron and neutral-hadron classes, as described in Section~\ref{sec:slddata}, while the SLD truth particles are assigned to all five classes.

The particle network learns the reconstructed-particle quantities $\pt$, $\eta$, $\phi$, $v_x$, $v_y$, $v_z$, and the particle class. The event network learns four global reconstructed quantities: $\HT$, reconstructed-particle multiplicity, $p_x^{\rm miss}$, and $p_y^{\rm miss}$. The event-level quantities are generated first and are used as conditioning information for the particle network.

The same particle-level preprocessing is used for ALEPH and SLD. Particles are restricted to $|\eta|\leq2.7$, with no particle-level \pt\ threshold. Transverse momentum is represented as $\pt^{\rm rel}=\pt/\HT$, $\phi$ is encoded as $(\sin\phi,\cos\phi)$, and the particle classes are represented as one-hot vectors.

\subsection{Architecture and training}

We use the two-network Parnassus flow-matching architecture described in Ref.~\cite{Dreyer:2025events}, with the same network configuration for ALEPH and SLD.

The particle network is a cross-attention transformer with eight dual-cross-attention blocks, hidden dimension 96, and six attention heads.  Adaptive layer normalization~\cite{Peebles:2022dit} conditions the network on the truth-event representation, global event information, and the flow time. The network generates the reconstructed-particle features using rectified conditional flow matching ~\cite{Lipman:2022fm,Liu:2022rf,Tong:2023cfm} and an MSE-plus-cosine-direction loss~\cite{Yao:2024fasterdit}.

The event network is a residual network with three residual blocks and hidden dimension 128. It generates $\HT$, reconstructed-particle multiplicity, $p_x^{\rm miss}$, and $p_y^{\rm miss}$. At inference, the event network is evaluated first and its generated quantities condition the particle network.

Both networks are trained for 300 epochs with batch size 128 using the Lion optimizer~\cite{Chen:2023lion} and a learning rate of $3\times10^{-5}$.  Stochastic weight averaging is applied during training.  Generation uses a DPM-Solver multistep integrator ~\cite{Lu:2022dpm} with 40 sampling steps. Further details of the Parnassus architecture and training procedure are given in Refs.~\cite{Dreyer:2024bhb,Dreyer:2025events,Elabd:2026parnassus}.

\subsection{Cross-detector fine-tuning}
\label{sec:protocol}

The transfer study focuses on the Parnassus particle network. Two SLD models are trained using the same architecture and the same SLD training and validation samples. The directly trained SLD model begins from a random initialization.  The fine-tuned model instead begins from the particle-network parameters learned on ALEPH.

The two initializations can be written as
\[
\theta_{\rm SLD}^{(0)} =
\begin{cases}
\theta_{\rm random},
& \text{directly trained SLD model}, \\[4pt]
\theta_{\rm ALEPH},
& \text{ALEPH-initialized SLD model}.
\end{cases}
\]

Only the learned particle-network parameters are transferred from ALEPH. The optimizer state, learning-rate schedule, training epoch, and stochastic-weight-averaging state are not carried over. Normalization constants are derived from the SLD training sample, so both SLD models use the same target-domain normalization.

After initialization, the two SLD models are trained under the same conditions.  The training and validation samples, preprocessing, normalization, architecture, optimizer, learning-rate schedule, loss function, batch size, and number of training epochs are held fixed. The comparison therefore isolates the effect of initializing the particle network with parameters learned from ALEPH.

For the main transfer result, the event network is trained directly on SLD and is shared by the two particle-network configurations. Its parameters are fixed when comparing the directly trained and ALEPH-initialized particle models. This removes changes in the event-level conditioning as a source of difference between the two particle models.

The ALEPH and SLD particle networks use the same architecture and feature representation, allowing the ALEPH parameters to initialize the SLD model. The reconstructed distributions themselves are different, as shown in Section~\ref{sec:domains}. Fine-tuning therefore does not assume that the ALEPH and SLD detector responses are the same. Instead, it tests whether parameters learned from the ALEPH detector response provide a better starting point for learning the SLD response than random initialization.

A separate control in which the event network is also initialized from ALEPH is discussed in Section~\ref{sec:levels}. The main comparison uses the common SLD event network so that the effect of particle-network transfer can be studied independently.

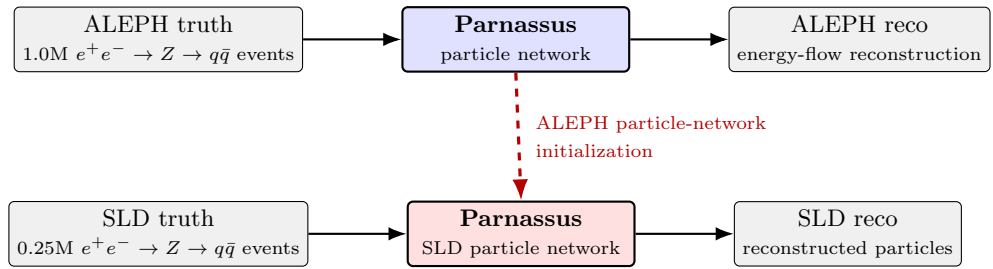
\begin{figure*}[t]
\centering
\begin{tikzpicture}[
  font=\small,
  box/.style={draw, rounded corners=2pt, minimum height=8.5mm,
    minimum width=30mm, align=center},
  net/.style={box, fill=blue!12, thick},
  netft/.style={box, fill=red!12, thick},
  dat/.style={box, fill=black!6},
  arr/.style={-{Latex[length=2.4mm]}, thick},
  xfer/.style={-{Latex[length=2.8mm]}, very thick, red!70!black, dashed},
]
\node[dat] (at) {ALEPH truth\\ \scriptsize 1.0M \eeZ\ events};
\node[net, right=13mm of at] (ap)
  {\textbf{Parnassus}\\
   \scriptsize particle network};
\node[dat, right=13mm of ap] (ar)
  {ALEPH reco\\
   \scriptsize energy-flow reconstruction};

\node[dat, below=17mm of at] (st)
  {SLD truth\\
   \scriptsize 0.25M \eeZ\ events};
\node[netft, right=13mm of st] (sp)
  {\textbf{Parnassus}\\
   \scriptsize SLD particle network};
\node[dat, right=13mm of sp] (sr)
  {SLD reco\\
   \scriptsize reconstructed particles};

\draw[arr] (at) -- (ap);
\draw[arr] (ap) -- (ar);
\draw[arr] (st) -- (sp);
\draw[arr] (sp) -- (sr);

\draw[xfer] (ap.south) -- node[right=1mm, align=left]
  {\scriptsize ALEPH particle-network\\
   \scriptsize initialization} (sp.north);
\end{tikzpicture}

\caption{Cross-detector fine-tuning from ALEPH to SLD. The ALEPH-trained particle-network parameters initialize the SLD particle network before training on the SLD sample. The directly trained SLD model uses the same architecture and training procedure with random initialization. The same SLD-trained event network is used for both particle models in the main comparison.}
\label{fig:schematic}
\end{figure*}

Figure~\ref{fig:schematic} summarizes the training procedure.

\section{Main results}
\label{sec:results}

\subsection{Evaluation metrics}
\label{sec:metrics}

All models are evaluated on the same fixed SLD test sample; 9{,}971 of its
9{,}972 events are retained after generation and postprocessing and are used
for all comparisons. Events are generated using 40 solver steps and then
processed with the same analysis procedure. Jets are clustered with the
anti-$k_t$ algorithm~\cite{Cacciari:2011ma} using $R=0.5$ and are required to
satisfy $\pt^{\rm jet}>20$\,GeV and to contain at least two constituents.

The present analysis retains the hadron-collider-inspired
coordinate and jet definitions used by the existing Parnassus analysis
pipeline. A future implementation specialized for $e^+e^-$ collisions could
instead represent particle directions using the polar angle, perform matching
using the three-dimensional opening angle, and reconstruct jets with an
exclusive $e^+e^-$ algorithm such as the Durham $k_t$
algorithm~\cite{Catani:1991hj}. Such choices would avoid privileging the beam
transverse plane and provide a representation more naturally adapted to the
approximately spherical geometry of $e^+e^-$ events.

Particle-level response is evaluated by matching charged truth and
reconstructed particles within each event. A Hungarian assignment is used to
minimize $\Delta R(\eta,\phi)$, and only matched pairs satisfying
$\Delta R<0.1$ are retained. The same matching procedure is applied to the
archived SLD reconstruction and to the generated samples, allowing the model
resolutions to be compared directly with the detector reference.

For a residual distribution $r$, we characterize the resolution using the central 68\% width,
\[
\sigma_{68}(r)
=
\frac{Q_{84}(r)-Q_{16}(r)}{2},
\]
where $Q_p$ denotes the $p$th percentile.

For the particle-level comparison, we consider the angular residual $\deta$ and the relative transverse-momentum residual $\dptr$. In addition to their $\sigma_{68}$ values, we report the fractions of matched particles satisfying $|\deta|<0.01$ and $|\dptr|<0.01$, as well as the matching efficiency.

Jet response is evaluated using nearest-jet matching with $\Delta R<0.3$. The same jet definition and matching procedure are used for the SLD reference and for each generated sample.

\subsection{Particle- and jet-level performance: direct training vs.\ fine-tuning}
\label{sec:headline}

Table~\ref{tab:headline} summarizes the main particle- and jet-level results.  For the angular resolutions, the table also gives the width relative to the SLD reference,
\[
\frac{\sigma_{68}^{\rm model}}{\sigma_{68}^{\rm SLD}}.
\]
A value of one indicates the same resolution width as the archived SLD reconstruction. Values larger than one correspond to a broader response.

\begin{table*}[t]
\centering
\caption{Particle- and jet-level performance on the fixed SLD test set. The SLD reference is the archived reconstruction evaluated against truth using the same matching definitions. Particle quantities use charged particle matching with $\Delta R<0.1$, while jet quantities use nearest-jet matching with $\Delta R<0.3$.}
\label{tab:headline}
\small
\begin{tabular}{lccc}
\toprule
 & SLD reference & Directly trained SLD & ALEPH$\to$SLD fine-tune \\
\midrule

\multicolumn{4}{l}{\it Charged particles} \\

Angular resolution $\sigma_{68}(\deta)$
    & 0.0022 & 0.0158 & \textbf{0.0031} \\

Width relative to SLD reference
    & 1.00$\times$ & 7.3$\times$ & \textbf{1.42$\times$} \\

Fraction with $|\deta|<0.01$
    & 91.5\% & 48.7\% & \textbf{92.0\%} \\

Relative-\pt\ resolution $\sigma_{68}(\dptr)$
    & 0.0167 & 0.0428 & \textbf{0.0175} \\

Fraction with $|\dptr|<0.01$
    & 49.4\% & 22.3\% & \textbf{50.1\%} \\

Matching efficiency
    & 0.783 & 0.672 & \textbf{0.756} \\

\midrule

\multicolumn{4}{l}{\it Jets} \\

Angular resolution $\sigma_{68}(\Delta\eta_{\rm jet})$
    & 0.0146 & 0.0253 & \textbf{0.0168} \\

Width relative to SLD reference
    & 1.00$\times$ & 1.73$\times$ & \textbf{1.15$\times$} \\

Azimuthal resolution $\sigma_{68}(\Delta\phi_{\rm jet})$
    & 0.0145 & 0.0258 & \textbf{0.0182} \\

Fraction with $|\Delta\eta_{\rm jet}|<0.01$
    & 54.4\% & 32.2\% & \textbf{48.7\%} \\

$D_2$ residual width $\sigma_{68}$
    & 0.185 & 0.301 & \textbf{0.226} \\

$C_2$ residual width $\sigma_{68}$
    & 0.026 & 0.041 & \textbf{0.032} \\

\bottomrule
\end{tabular}
\end{table*}

At the particle level, the directly trained SLD model gives $\sigma_{68}(\deta)=0.0158$, compared with 0.0022 for the SLD reference. Its angular-response width is therefore $7.3\times$ the reference value. After initialization from the ALEPH model, the width decreases to 0.0031, or $1.42\times$ the SLD reference.  The fraction of matched particles satisfying $|\deta|<0.01$ increases from 48.7\% to 92.0\%, comparable to the reference value of 91.5\%.

The relative-\pt\ response also improves after fine-tuning. Its $\sigma_{68}$ decreases from 0.0428 to 0.0175, compared with 0.0167 for the SLD reference. The fraction of matched particles satisfying $|\dptr|<0.01$ increases from 22.3\% to 50.1\%, while the reference value is 49.4\%.  The matching efficiency increases from 0.672 to 0.756, compared with 0.783 for the SLD reconstruction.

\begin{figure*}[t]
\centering
\includegraphics[width=\textwidth]{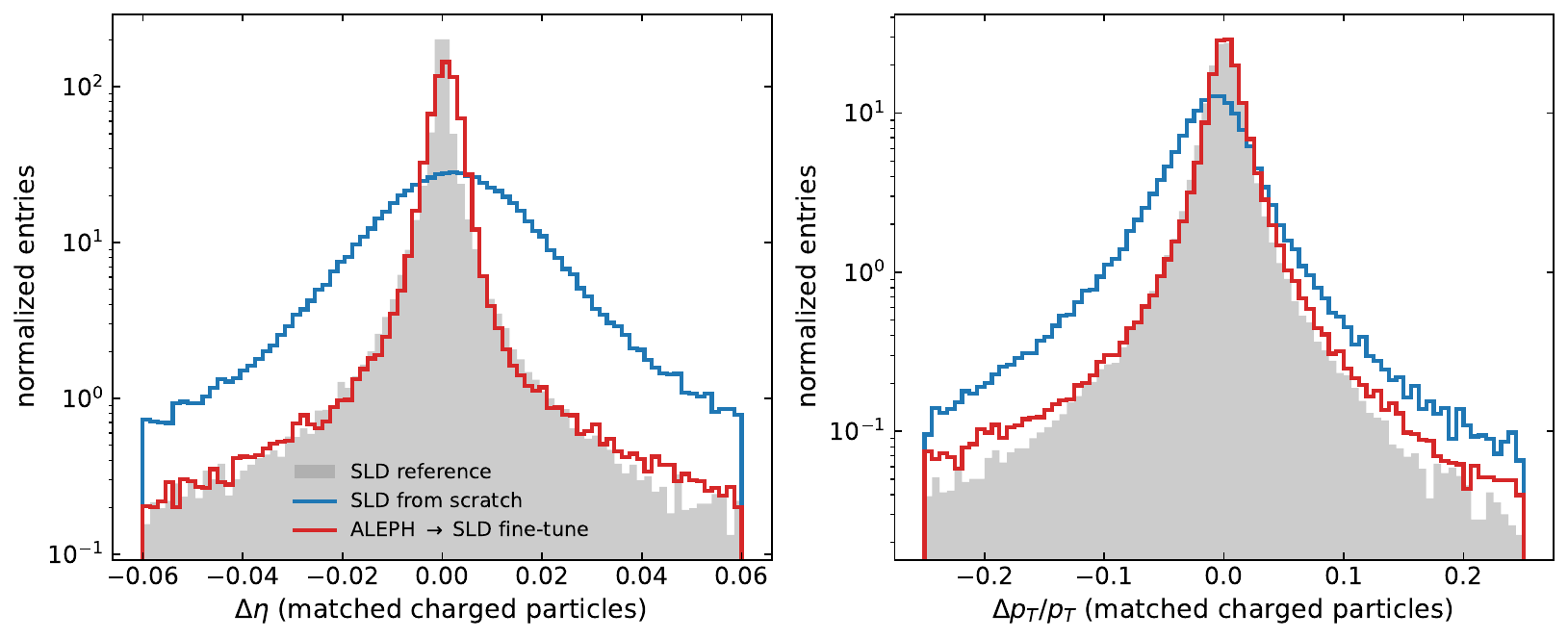}
\caption{Residual distributions for matched charged particles in the SLD reference, the directly trained SLD model, and the ALEPH$\to$SLD fine-tuned model. The left panel shows \deta\ and the right panel shows \dptr.  Particle pairs are matched with $\Delta R<0.1$.}
\label{fig:particle3way}
\end{figure*}

Figure~\ref{fig:particle3way} shows the corresponding charged-particle residual distributions.  The directly trained model has a substantially broader angular response than the SLD reference. The fine-tuned model recovers the narrow central structure of the reference distribution. The improvement in the momentum response is also visible, although it is smaller than that observed for the angular response.

The improvement remains visible after the generated particles are clustered into jets.  For the jet $\eta$ response, the directly trained model has a width of 0.0253, corresponding to $1.73\times$ the SLD reference width of 0.0146. The fine-tuned model gives a width of 0.0168, or $1.15\times$ the reference. The jet $\phi$ resolution also moves closer to the SLD reference after fine-tuning.  Similar improvements are observed in the $D_2$ and $C_2$ residual widths.

The improvement from ALEPH initialization therefore persists after the generated particles are clustered into jets.

\subsection{Weight changes during cross-detector fine-tuning}
\label{sec:drift}

To examine how the ALEPH-pretrained particle network changes during fine-tuning on SLD, we compare the parameters of each network module with their values in the ALEPH model. For a module with parameters $\theta$, we define the relative parameter difference as
\[
d(\theta)
=
\frac{\|\theta-\theta_{\rm ALEPH}\|}
     {\|\theta_{\rm ALEPH}\|}.
\]
The same quantity is evaluated for the directly trained SLD model. Since the directly trained model has the same architecture but is initialized independently, its distance from the ALEPH model provides a reference for the difference between independently trained networks.

\begin{figure*}[t]
\centering
\includegraphics[width=0.62\textwidth]{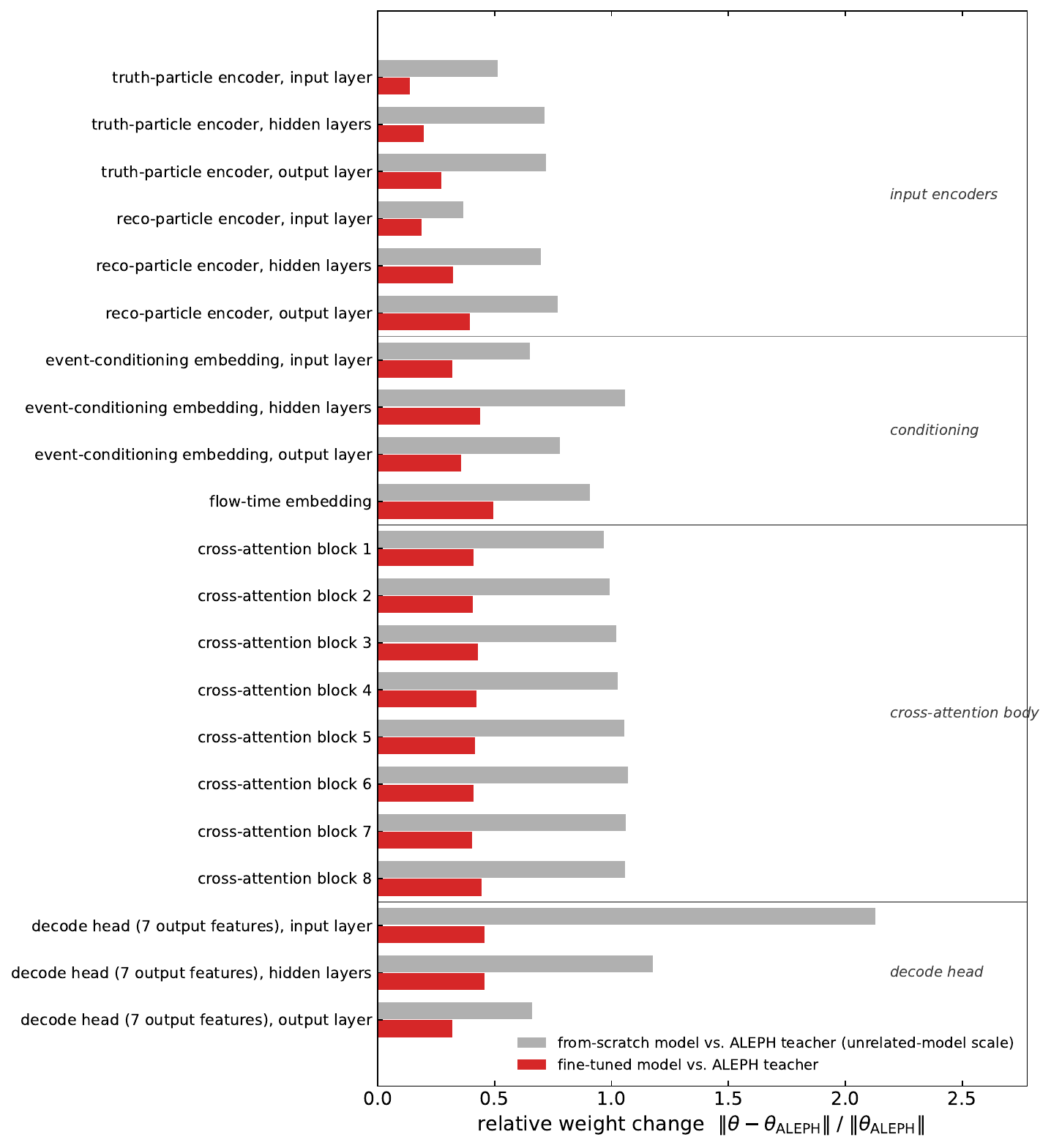}
\caption{Relative parameter difference with respect to the pretrained ALEPH particle network for the main module groups. The modules are ordered following the particle-network forward pass from the input encoders to the decode head.  The ALEPH$\to$SLD fine-tuned model is compared with the directly trained SLD model, which has the same architecture but independent initialization. Smaller values indicate parameters that remain closer to the ALEPH model.}
\label{fig:drift}
\end{figure*}

Figure~\ref{fig:drift} shows the parameter differences in the order in which the main modules enter the particle-network forward pass. The first group contains the truth-particle and flow-state encoders, which map the 12-dimensional particle representation to the 96-dimensional latent representation used by the network.  The truth-particle encoder shows the smallest changes during fine-tuning, while somewhat larger changes are observed in the flow-state encoder.

The next group contains the event-conditioning and flow-time embeddings. These modules provide the conditioning information used throughout the particle network. Their parameter differences are larger than those of the truth-particle encoder.

The eight cross-attention blocks form the main body of the particle network.  They exchange information between the current flow-state particle cloud and the truth-particle cloud and are used to construct the latent representation from which the flow velocity is predicted. The relative parameter differences are similar across the eight blocks, with no strong progression from the first to the last block.

The final group is the decode head, which combines the final 96-dimensional particle representation with the 288-dimensional event context and maps it to the 12-dimensional flow velocity in the encoded particle space. The input and hidden layers of the decode head show changes comparable to those in the attention body, while the output layer changes somewhat less.

For every module group shown in Fig.~\ref{fig:drift}, the fine-tuned model remains closer to the ALEPH parameters than the directly trained SLD model. No particle-network parameters are frozen during fine-tuning, so this proximity is not imposed by the training procedure. Although all particle-network parameters are trainable, the fine-tuned model remains closest to the ALEPH model in the particle encoders, with larger changes in the conditioning, attention, and decoding components.

\subsection{Event-, jet-, and particle-level performance}
\label{sec:levels}

\begin{figure*}[t]
\centering
\includegraphics[width=\textwidth]{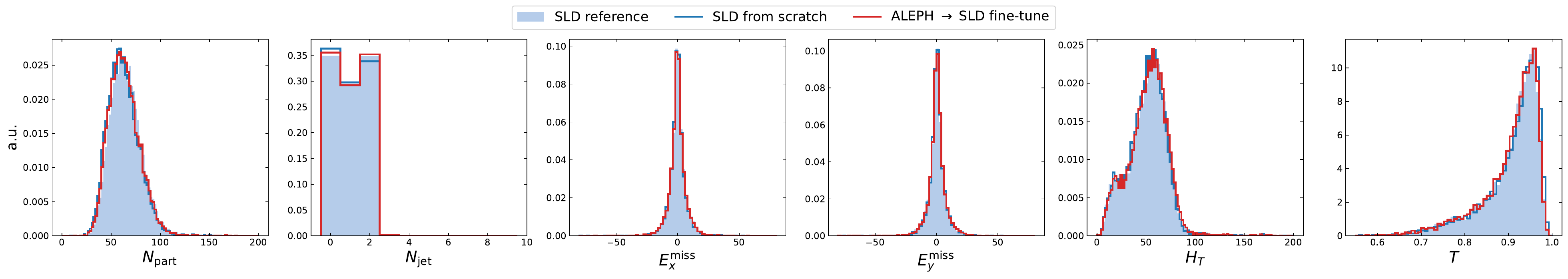}
\caption{Event-level distributions on the fixed SLD test set for the archived SLD reference, the directly trained SLD model, and the ALEPH$\to$SLD fine-tuned model. The observables include reconstructed-particle multiplicity, jet multiplicity, missing momentum, \HT, and thrust.}
\label{fig:exteventfalse}
\end{figure*}

\begin{figure*}[t]
\centering
\includegraphics[width=\textwidth]{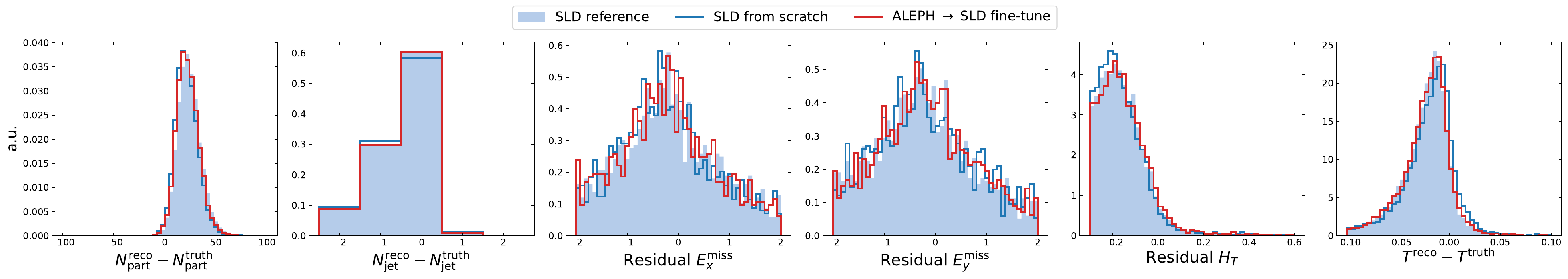}
\caption{Event-level residual distributions on the fixed SLD test set, including the residuals of reconstructed-particle multiplicity, jet multiplicity, missing momentum, \HT, and thrust.}
\label{fig:exteventtrue}
\end{figure*}

\begin{figure*}[t]
\centering
\includegraphics[width=\textwidth]{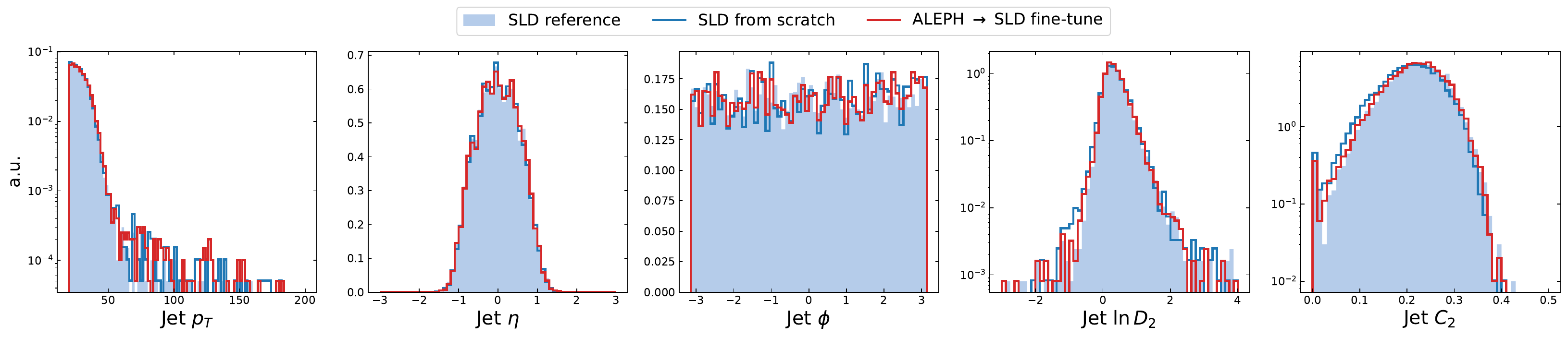}
\caption{Jet-level distributions on the fixed SLD test set for the archived SLD reference, the directly trained SLD model, and the ALEPH$\to$SLD fine-tuned model. The comparison includes the $\ln D_2$ and $C_2$ substructure observables.}
\label{fig:extjetfalse}
\end{figure*}

\begin{figure*}[t]
\centering
\includegraphics[width=\textwidth]{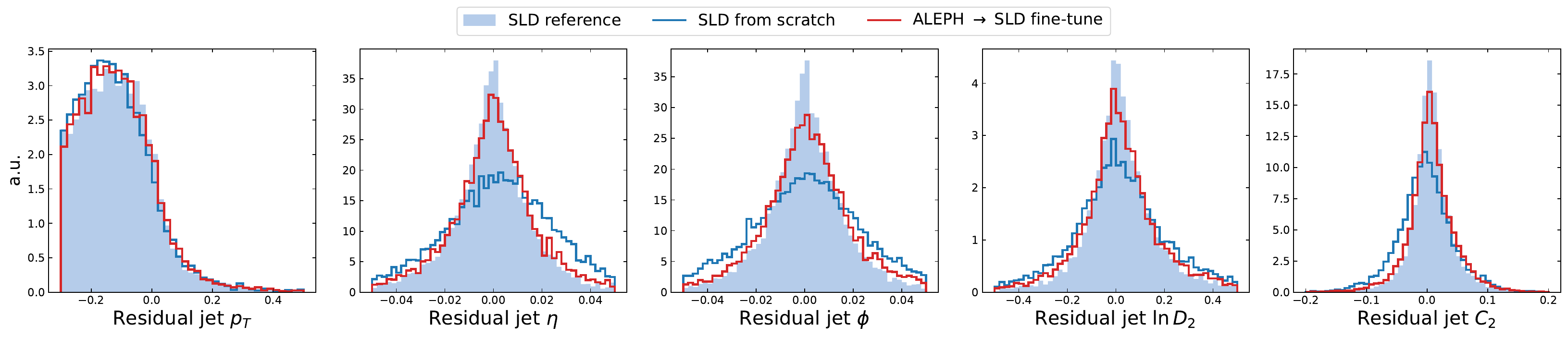}
\caption{Jet-level residual distributions on the fixed SLD test set. Jets are matched using nearest-jet matching with $\Delta R<0.3$.}
\label{fig:extjettrue}
\end{figure*}

\begin{figure*}[t]
\centering
\includegraphics[width=\textwidth]{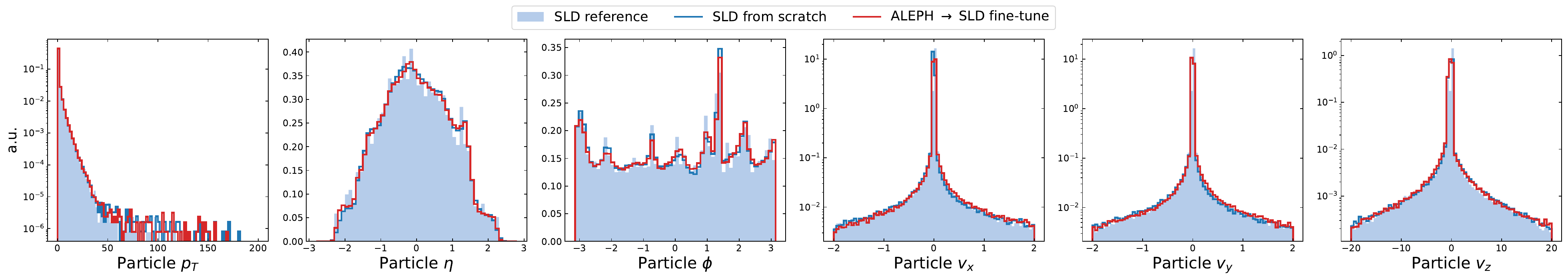}
\caption{Reconstructed-particle distributions on the fixed SLD test set for the archived SLD reference, the directly trained SLD model, and the ALEPH$\to$SLD fine-tuned model.}
\label{fig:extpartfalse}
\end{figure*}

\begin{figure*}[t]
\centering
\includegraphics[width=\textwidth]{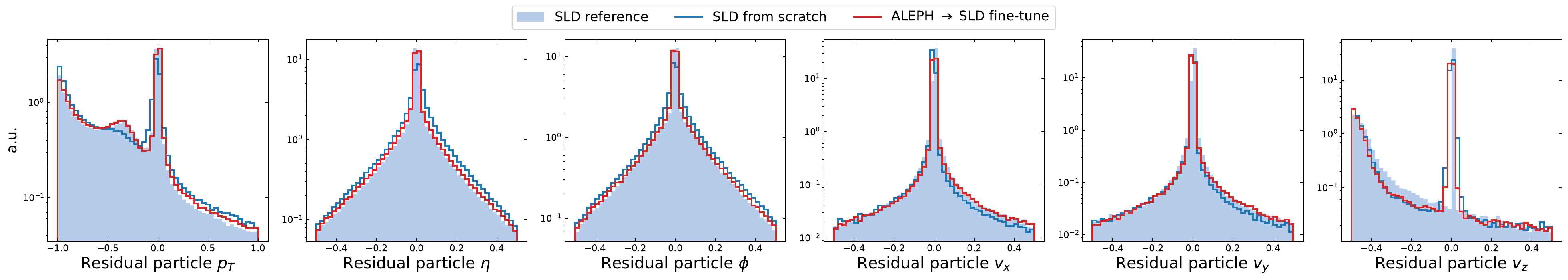}
\caption{Particle-level residual distributions on the fixed SLD test set using the standard postprocessing matching employed for the full comparison plots.  Quantitative particle-resolution measurements use the matching definition of Section~\ref{sec:metrics}.}
\label{fig:extparttrue}
\end{figure*}

Figures~\ref{fig:exteventfalse}--\ref{fig:extparttrue} present the event-, jet-, and particle-level comparison on the fixed SLD test set. For each level, marginal distributions are shown first, followed by the corresponding residual distributions. The event-level comparison also includes the thrust $T$, calculated from the reconstructed particles using the same procedure as in the ALEPH study~\cite{Lo:2026aleph}.

\paragraph{Event level.}
Figure~\ref{fig:exteventfalse} compares the reconstructed-particle multiplicity, jet multiplicity, missing-momentum components, \HT, and thrust. The directly trained SLD model and the ALEPH$\to$SLD fine-tuned model both reproduce the main features of the SLD reference distributions. The corresponding residual distributions in Fig.~\ref{fig:exteventtrue} are also similar for the two models, including those for \HT, missing momentum, and thrust.

As an additional check, we trained a model in which the event network, in addition to the particle network, was initialized from the corresponding ALEPH model. This produced negligible changes in the particle-level results and only small changes at the jet level. We therefore do not observe a significant additional benefit from cross-detector initialization of the event network in this study.

\paragraph{Jet level.}
Figures~\ref{fig:extjetfalse} and~\ref{fig:extjettrue} show the corresponding comparison for jets reconstructed from the generated particles. Both models describe the jet $\pt$, $\eta$, and $\phi$ distributions, as well as the $\ln D_2$ and $C_2$ substructure observables. The residual distributions provide a clearer distinction between the two models. The fine-tuned model gives narrower jet angular residuals and reproduces the $D_2$ and $C_2$ residual distributions more closely than the directly trained SLD model. These differences are consistent with the quantitative jet-response comparison in Table~\ref{tab:headline}.

\paragraph{Particle level.}
The reconstructed-particle distributions are shown in
Fig.~\ref{fig:extpartfalse}. Both models reproduce the main features of the SLD reference distributions in $\pt$, $\eta$, $\phi$, and the reconstructed vertex coordinates. In particular, both models describe the nonuniform structure of the SLD $\phi$ distribution.

The particle residuals in Fig.~\ref{fig:extparttrue} use the standard postprocessing matching employed for the full comparison plots. With this convention, the directly trained SLD model gives broader angular residual distributions, while the ALEPH$\to$SLD fine-tuned model produces a narrower response that is closer to the SLD reference. The directly trained SLD model gives broader angular residual distributions, while the ALEPH$\to$SLD fine-tuned model produces a narrower response that is closer to the SLD reference. An improvement is also visible in the particle momentum response.

\subsection{Flavour dependence of the reconstructed $Z$ mass}
\label{sec:flavour}

\begin{figure*}[t]
\centering
\includegraphics[width=0.86\textwidth]{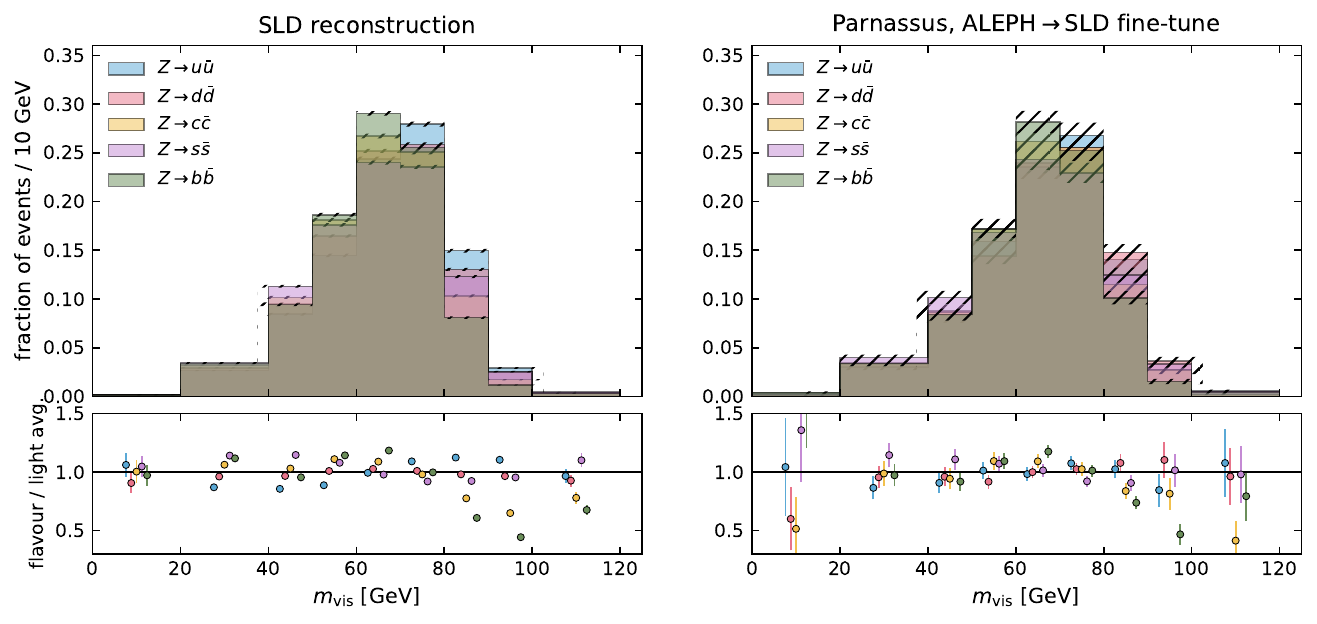}
\caption{Reconstructed visible mass split by the flavour of the primary quark pair, for the archived SLD reconstruction on the full simulated sample (left) and for the ALEPH$\to$SLD fine-tuned model on the test sample (right). Distributions are normalized to unit area, and the lower panels show each flavour divided by the light-flavour ($u$, $d$, $s$) average. Hatching indicates the statistical uncertainty.}
\label{fig:flavour}
\end{figure*}

As an additional check of the conditional detector response, we study the reconstructed visible mass separately for different primary $Z\to q\bar q$ flavours. Figure~\ref{fig:flavour} shows that the archived SLD reconstruction has a clear flavour dependence, with heavy-flavour events shifted toward lower visible mass relative to light flavours. The fine-tuned Parnassus model reproduces this overall pattern well, although some discrepancy remains for charm.

The remaining difference is mainly associated with the neutral-particle response. In the current representation, all neutral hadrons are grouped into a single particle class, while the SLD reconstruction has different responses to different neutral species. A more detailed neutral-particle classification may therefore further improve the flavour-dependent response.

\section{Conclusions}
\label{sec:conclusions}

We have demonstrated cross-detector transfer learning for generative detector simulation using Parnassus. A particle network pretrained on 1.0M archived ALEPH hadronic $Z$ events was fine-tuned on the 250k-event SLD sample and compared with an identical network trained directly on SLD. The only difference between the two SLD training configurations is the initialization of the particle-network parameters. Cross-detector fine-tuning improves the charged-particle angular resolution from $7.3\times$ to $1.42\times$ the SLD reference width and the corresponding jet angular resolution from $1.73\times$ to $1.15\times$.  Improvements are also observed in the particle momentum response, matching efficiency, and jet substructure observables. The fine-tuned network remains closer to the pretrained ALEPH model than the directly trained SLD network across all major particle-network components, consistent with partial retention of the ALEPH-trained parameters during adaptation to SLD.

A generative detector-simulation model trained for one experiment can therefore provide a useful initialization for another detector with related underlying physics. Cross-detector pretraining may therefore provide a practical way to
reuse preserved simulated samples when developing generative models for legacy collider datasets~\cite{Cheng:2026sld,Defranchis:2026aleph}. The
ALEPH$\to$SLD study presented here provides a first step toward such
cross-detector pretrained generative-simulation models.

\section{DATA AND CODE AVAILABILITY}
The archived SLD Monte Carlo dataset used in this study is available on Zenodo at \url{https://doi.org/10.5281/zenodo.22732257}. The code used for training is available at \url{https://github.com/parnassus-hep/cms-flow-evt}.

\begin{acknowledgments}
We are grateful to the SLAC accelerator staff that operated SLC and the SLD Collaboration for collecting and curating this unique and scientifically rich dataset. These activities were supported by the U.S. Department of Energy (DOE) and the U.S. National Science Foundation; the Istituto Nazionale di Fisica Nucleare of Italy; the Japan–US Cooperative Research Project on High Energy Physics; and, in the United Kingdom, the Science and Engineering Research Council and its successor the Particle Physics and Astronomy Research Council.

We thank the SLD Collaboration for allowing us to resurrect and release their data and documentation; in particular, we are indebted to Martin Breidenbach, Tony Johnson, and Su Dong for extensive conversations and for loaning us artifacts related to these data.

We thank Dmitrii Kobylianskii and Eilam Gross for the technical support with Parnassus. 

We also appreciate Luna Chen and our other colleagues in the electron-positron alliance for useful conversations.

 This work is supported by the U.S. Department of Energy (DOE) under Contract No. DE-AC02-76SF00515 and by the CompHEP Center for Computational Excellence.  In particular, this research was catalyzed and carried out by the Genesis Mission, in part, through the HEP pilot project Knowledge Extraction. We are grateful to Katrin Heitmann and all of our other collaborators on this project.
 
 This research used resources of the National Energy Research Scientific Computing Center (NERSC), a DOE Office of Science User Facility supported by the Office of Science of the U.S. Department of Energy under Contract No.~DE-AC02-05CH11231, under NERSC award HEP-ERCAP0035546.
\end{acknowledgments}

\bibliographystyle{apsrev4-2}
\clearpage
\bibliography{references}

\end{document}